\documentclass[11pt,a4,twoside]{article}
\usepackage[dvips,a4paper,left=2.5cm,right=2.5cm,top=2.5cm,bottom=2.5cm,headsep=1em]{geometry}
\usepackage{fancyhdr}
\usepackage{titlesec}
\usepackage[latin1]{inputenc}
\usepackage{amsmath,amsfonts,amssymb,array}
\usepackage{rotating}
\usepackage[font={small}]{caption}
\usepackage{natbib}
\usepackage{lmodern,slantsc}
\usepackage{multirow}
\usepackage{chngcntr}
\usepackage{placeins}
\usepackage{caption}
\usepackage{enumitem}
\usepackage[breaklinks,
colorlinks=true, linkcolor=blue, citecolor=black, urlcolor=blue,
pdfborder={0 0 0}, pdfpagelabels]{hyperref}
\def\vx{\mathbf x}

\def\vv{\mathbf v}

\def\vC{\mathbf C}

\def\vU{\mathbf U}

\def\v0{\boldsymbol{0}}

\def\vgamma{\boldsymbol{\gamma}}

\def\vlambda{\boldsymbol{\lambda}}

\def\L{\langle}
\def\R{\rangle}
\newlength{\FigureHeight}
\newlength{\FigureHeightHalf}

\numberwithin{equation}{section}
\titleformat{\section}
{\large\bfseries}{\thetitle.}{0.5em}{}
\titleformat{\subsection}
{\normalfont\itshape\filcenter}{\normalfont\thetitle.}{0.5em}{}
\makeatletter
\def\@fnsymbol#1{\ensuremath{\ifcase#1\or *\or \dagger\or \ddagger\or
   \mathsection\or \mathparagraph\or \|\or **\or \dagger\dagger
   \or \ddagger\ddagger \else\@ctrerr\fi}}
\makeatother

\begin{document}

\title{\vspace{0.0em} Comment on `Lie symmetry analysis of the Lundgren-Monin-Novikov equations for multi-point\\ probability density functions of turbulent flow'}
\author{M. Frewer$\,^1$\thanks{Email address for correspondence:
frewer.science@gmail.com}$\:\,$, G. Khujadze$\,^2$ \& H. Foysi$\,^2$\\ \\
\small $^1$ Heidelberg, Germany\\
\small $^2$ Chair of Fluid Mechanics, Universit\"at Siegen, 57068
Siegen, Germany}
\date{{\small\today}}
\clearpage \maketitle \thispagestyle{empty}

\vspace{-2.0em}\begin{abstract}

\noindent The recent study by Wac{\l}awczyk~{\it et~al.}~[\href{http://iopscience.iop.org/article/10.1088/1751-8121/aa62f4/meta}{J.~Phys.~A:~Math.~Theor.$\,$50,~175501~(2017)}] possesses\linebreak three shortcomings: (i)~The analysis misses a key aspect of the LMN equations which makes their Lie-group symmetry results incomplete. In particular, two essential symmetries will break when including this aspect. (ii)~The statements on the constraints regarding the infinite-dimensional symmetry groups are misleading. (iii)~The particular symmetries originating solely from the linearity of the LMN hierarchy violate the classical principle of cause and effect and therefore are unphysical. Within this Comment we present a detailed proof to this claim and conclude with the note that the new study by Wac{\l}awczyk~{\it et~al.} gives an unrealistic outlook on deriving invariant symmetry solutions for velocity correlations that arise from intermittent processes.

\vspace{0.5em}\noindent{\footnotesize{\bf Keywords:} {\it Statistical Physics, Turbulence, Probability Density Functions, Lie Groups, Statistical Symmetries, Symmetry Breaking, Principle of
Causality, Closure Problem, Scaling Laws}}
\end{abstract}

\section{Incompleteness of symmetry analysis \label{Sec1}}
The Lundgren-Monin-Novikov (LMN) equations are accompanied by at least five physical constraints in order to guarantee their solutions to be physical. These are the four well-known and so-called non-negativity, normalization, coincidence and separation constraints,\footnote[2]{Cf., e.g., \cite{Lundgren67,Hosokawa06,Friedrich12}.} and a fifth one, not so well-known but an equally important constraint, the conditional constraint first derived by \cite{Ievlev70} and then later also discussed by \cite{Monin75}.\footnote[3]{For the incompressible case, see Eq.~[2.6] in \cite{Ievlev70}, and Eq.~[19.139] in \cite{Monin75}.}

Although these constraints (up to the Ievlev-constraint) get mentioned in \cite{Oberlack17}, they are, except for the normalization constraint (Eq.~[10]), not included into their symmetry analysis.
In principle, it is this approach which renders their symmetry analysis incomplete, because these constraints are part of the system itself defining the LMN equations. They definitely may not be treated as exogenous conditions like, for example, initial or boundary conditions which under certain asymptotic assumptions can be neglected or ignored within the search for symmetries of an underlying dynamical equation.~In particular, due to not including all these internal LMN constraints into their symmetry analysis in a strict systematic manner, it has been overlooked in \cite{Oberlack17} that the two ``new" symmetries $X_*$ (Eq.~[47]) and $X_{**}$~(Eq.~[48]) are both broken symmetries. It is straightforward to show that both these symmetries are not compatible to the separation constraint (here only shown for the two-point probability density function (PDF) $f_2$)
\begin{equation}
\lim_{|\vx_{(2)}-\vx_{(1)}|\rightarrow\infty}f_2(\vv_{(2)},\vx_{(2)},\vv_{(1)},\vx_{(1)},t)=f_1(\vv_{(2)},\vx_{(2)},t)\cdot f_1(\vv_{(1)},\vx_{(1)},t),\label{170830:1149}
\end{equation}
which expresses the natural property of statistical independence when two points are infinitely far apart. A crucial property which now is not maintained when transforming the variables according to
the proposed symmetry $X_*$ (Eq.~[47]), or to  $X_{**}$~(Eq.~[48]), which in finite (global) form will read, respectively,\footnote[2]{To note is that the functions $b_{(n)}^0$
for $X_*$ (Eq.~[47]) are different to those of $X_{**}$~(Eq.~[48]), simply due to the fact that they are constrained differently by the LMN normalization constraint. In this regard, it is also to be noted that $X_{*}$
is not commuting with $X_{**}$, as mistakenly presented in Tab.~[1] in \cite{Oberlack17}.}
\begin{align}
\text{T}_{X_{*}}\!:&\quad\; t^*=t,\quad \vx_{(i)}^*=\vx_{(i)},\quad \vv^*_{(i)}=\vv_{(i)},\quad f^*_n=e^{a_s}f_n+(1-e^{a_s})b_{(n)}^0,\label{170830:1150}\\[0.5em]
\text{T}_{X_{**}}\!:&\quad\; t^*=t,\quad \vx_{(i)}^*=\vx_{(i)},\quad \vv^*_{(i)}=\vv_{(i)},\quad f^*_n=f_n+b_{(n)}^0,\label{170901:1745}
\end{align}
where $a_s$ is a group parameter of the infinitesimal $X_{*}$, constrained according to the LMN constraints (see Sec.~[4] in \cite{Oberlack17}). Instead of the invariant result\footnote[3]{The abbreviation $n^*$ is used to symbolize the independent variable set $(\vv^*_{(n)},\vx^*_{(n)},t^*)$.}
\begin{equation}
\lim_{|\vx^*_{(2)}-\vx^*_{(1)}|\rightarrow\infty}f^*_2(2^*,1^*)=f^*_1(2^*)\cdot f^*_1(1^*),\label{170830:1314}
\end{equation}
one obtains the non-invariant result
\begin{align}
\!\!\lim_{|\vx^*_{(2)}-\vx^*_{(1)}|\rightarrow\infty}f^*_2(2^*,1^*)=&\, e^{-a_s}f^*_1(2^*)f^*_1(1^*)+(1-e^{-a_s})\Big(b_{(1)}^0(2^*)f^*_1(1^*)+b_{(1)}^0(1^*)f^*_1(2^*)\Big)
\nonumber\\[-0.5em]
&\!\!+e^{a_s}(1\!-e^{-a_s})^2\, b_{(1)}^0(2^*)b_{(1)}^0(1^*)+\! \lim_{|\vx^*_{(2)}-\vx^*_{(1)}|\rightarrow\infty}\!(1\!-e^{a_s})b^0_{(2)}(2^*,1^*),\label{170830:1237}
\end{align}
when transforming the separation constraint \eqref{170830:1149} according to the symmetry transformation $\text{T}_{X_{*}}$~\eqref{170830:1150}, and the following non-invariant result
\begin{align}
\lim_{|\vx^*_{(2)}-\vx^*_{(1)}|\rightarrow\infty}f^*_2(2^*,1^*)=&\, f^*_1(2^*)f^*_1(1^*)-b_{(1)}^0(2^*)f^*_1(1^*)-b_{(1)}^0(1^*)f^*_1(2^*)\nonumber\\[-0.5em]
& + b_{(1)}^0(2^*)b_{(1)}^0(1^*)+\lim_{|\vx^*_{(2)}-\vx^*_{(1)}|\rightarrow\infty} b^0_{(2)}(2^*,1^*),\label{170901:1758}
\end{align}
when transforming it according to $\text{T}_{X_{**}}$~\eqref{170901:1745}. Now obviously, since the $b_{(n)}^0$ by definition and construction do not dependent on the PDFs $f_n$, neither the functional structure of relation \eqref{170830:1237}, nor that of \eqref{170901:1758} can be reduced to the invariant result \eqref{170830:1314} for all admissible $f_n$ when mapped to $f_n^*$, irrespective of how the functions~$b_{(n)}^0$ are restricted. Hence, the separation constraint \eqref{170830:1149} does not stay invariant under $\text{T}_{X_*}$ \eqref{170830:1150} and $\text{T}_{X_{**}}$~\eqref{170901:1745}, with the effect that physical solutions get mapped into unphysical ones. A~mapping between physical solutions can only be achieved if $a_s=0$ and $b_{(n)}^0=0$, i.e, if the symmetries $\text{T}_{X_*}$ \eqref{170830:1150} and $\text{T}_{X_{**}}$~\eqref{170901:1745} are broken.

This breaking of both symmetries $\text{T}_{X_*}$ \eqref{170830:1150} and $\text{T}_{X_{**}}$~\eqref{170901:1745} has already been discussed in \cite{Frewer15}, and in more detail in \cite{Frewer14.1}, where the functions $b_{(n)}^0$ took the special form of Dirac-delta distributions. To note is that both transformations $\text{T}_{X_*}$~\eqref{170830:1150} and $\text{T}_{X_{**}}$~\eqref{170901:1745} are admitted as symmetries for all $n\geq 1$ by the infinite hierarchy of LMN equations (when excluding the separation constraints). This has first been derived and reported in \cite{Oberlack14}, where for simplicity the functions $b_{(n)}^0$ were specified to Dirac-delta~distributions.

Despite the fact shown here, that the separation constraint \eqref{170830:1149} breaks the symmetries $\text{T}_{X_*}$~\eqref{170830:1150} and $\text{T}_{X_{**}}$~\eqref{170901:1745} in a non-approximative manner, these symmetries are unphysical {\it per~se}. As shown in Sec.~\ref{Sec3}, both symmetries violate the causality principle of classical mechanics. Hence, these symmetries are to be discarded, otherwise they will unnecessarily lead to misleading results in turbulence research, as already several times demonstrated, e.g., in \cite{Frewer14.1,Frewer14.2} and \cite{Frewer16.1}.

\newgeometry{left=2.5cm,right=2.5cm,top=2.5cm,bottom=1.75cm,headsep=1em}

\section{Non-utilizable translation symmetry \label{Sec2}}

\vspace{-0.25em}
Not explicitly referred to as such, the infinite-dim\-en\-sio\-nal symmetry group $X_{**}$ (Eq.~[48]) is nothing else but the superposition principle of the unclosed first order equation (Eq.~[3]) of the linear LMN hierarchy. By definition, a superposition symmetry is built up by translation operators in the dependent variables with coefficients being solutions of the considered linear system of equations. Indeed, the symmetry $X_{**}$,
when formulated in its finite (global) form $\text{T}_{X_{**}}$~\eqref{170901:1745} is exactly of this translational type, where the functions $b_{(n)}^0$ are solutions of the LMN equations, linearly superposed on some PDF solution $f_n$ to yield a new solution $f_n^*$. Of course, all these functions will be constrained accordingly when also including the internal physical constraints of the LMN equations as listed in the beginning of Sec.~\ref{Sec1}.

For unclosed systems as the LMN equations, however, such a symmetry as $\text{T}_{X_{**}}$~\eqref{170901:1745} cannot be utilized in a promising, successful way, since it is not clear how to generate solutions without modelling the system. Hence the result in \cite{Oberlack17}, that the coefficients $b_{(n)}^0$ of the infinitesimal symmetry $X_{**}$ must be ``solutions" of the unclosed Eq.~[3], is misleading.
Obviously, to guess a solution for the lowest order equation is not the method of choice, since there is no guarantee {\it a priori} that, firstly, this guess is consistent also for all higher order equations and, secondly, that this guess in the end represents a real physical solution matching the direct numerical simulation (DNS) data. Hence, in our opinion, the symmetry $X_{**}$ is of no practical value for further investigations, and surely will also break as soon as the truncated Eq.~[3] is modelled, since any appropriate model will definitely be non-linear, thus not sharing anymore the general property of linear superposition.
\vspace{-0.5em}

\section{Violation of causality principle \label{Sec3}}

\vspace{-0.25em}
Even when ignoring the fact of Sec.~\ref{Sec1}, that of internal symmetry breaking, the statistical symmetries $\text{T}_{X_*}$~\eqref{170830:1150} and $\text{T}_{X_{**}}$~\eqref{170901:1745} are unphysical {\it per se} in violating the causality principle of classical mechanics.\footnote[2]{For more details on the particular causal structure we address here, please see Appendix \ref{SecA}.}
These symmetries, which have their origin solely from the linear structure of\linebreak[4] the LMN equations, clearly lead to wrong and misleading conclusions in turbulence, in particular when used further to generate statistical scaling laws. This fact has been proven analytically in a rigorous manner and demonstrated numerically several times by comparing to DNS data for different flow configurations. A strong mismatch between theory and (numerical) experiment is constantly observed when including both or one of these symmetries $\text{T}_{X_*}$~\eqref{170830:1150} and $\text{T}_{X_{**}}$~\eqref{170901:1745} or any of its variants into the analysis \citep{Frewer14.2,Frewer16.1}.

The analytic proof that \eqref{170830:1150} and \eqref{170901:1745} or any of its variants violate the classical principle of causality has already been given both for the system of PDFs as well as for its induced system of (multi-point) velocity moments \citep{Frewer15, Frewer14.1,Frewer14.2} --- for a general discussion on the physical aspect of statistical symmetries emerging from a dynamical system, please also see \cite{Frewer16.2}. Here we repeat again the analytic proof for the velocity moments, however, now in a more general setting by allowing the fluctuating velocities not only to scale deterministically but also randomly.

Let us first focus on the combined scaling and translation symmetry $\text{T}_{X_*}$~\eqref{170830:1150}, which is admitted as symmetry for all $n\geq 1$ by the infinite hierarchy of LMN equations (when {\it excluding} the separation constraints). In order to grasp the essence of the full proof, let us in a first step first consider the specific case where the translation functions $b_{(n)}^0$ are the following simple distributional LMN solutions
\citep{Oberlack14,Frewer15}\vspace{-0.4em}
\begin{equation}
b_{(n)}^0=\delta(\vv_{(1)})\cdots\delta(\vv_{(n)}),\label{170901:2143}
\vspace{-0.4em}
\end{equation}
in order to specify the general symmetry $\text{T}_{X_*}$~\eqref{170830:1150} first as\vspace{-0.4em}
\begin{equation}
\text{T}^\delta_{X_{*}}\!:\quad\; t^*=t,\quad \vx_{(i)}^*=\vx_{(i)},\quad \vv^*_{(i)}=\vv_{(i)},\quad f^*_n=e^{a_s}f_n+(1-e^{a_s})\delta(\vv_{(1)})\cdots\delta(\vv_{(n)}),\label{170901:2134}
\vspace{-0.4em}
\end{equation}
to then, in a second step, lift it back to the general case $\text{T}_{X_*}$~\eqref{170830:1150}.\footnote[3]{To note is that the specification \eqref{170901:2143} satisfies all conditions associated to the corresponding infinitesimal $X_*$~(Eq.~[47]) resulting from the LMN normalization constraint for \eqref{170901:2134}: $\int d\vv_{(1)} b_{(1)}^0=1$, $\int d\vv_{(2)} b_{(2)}^0=b_{(1)}^0$, etc.} Obviously, $\text{T}^\delta_{X_{*}}$~\eqref{170901:2134} induces\pagebreak[4] 

\restoregeometry

\noindent
the following symmetry transformation \citep{Oberlack15, Oberlack14}
\begin{equation}
\text{S}^\delta_{X_*}\!:\quad\; t^*=t,\quad \vx_{(i)}^*=\vx_{(i)},\quad \L \vU_{(1)}\otimes\cdots\otimes \vU_{(n)}\R^*=e^{a_s}\, \L \vU_{(1)}\otimes\cdots\otimes \vU_{(n)}\R,\;\;\; \forall n\geq 1,\label{170830:2045}
\end{equation}
for the multi-point velocity moments (Eq.~[1])
\begin{equation}
\L \vU_{(1)}\otimes\cdots\otimes \vU_{(n)}\R=\int d\vv_{(1)}\cdots d\vv_{(n)}\, f_n\cdot \vv_{(1)}\otimes\cdots\otimes \vv_{(n)},
\end{equation}
where $\L\cdot \R$ is the ensemble average operator, and $\vU_{(n)}=\vU(\vx_{(n)},t)$ the full instantaneous (not Reynolds-decomposed) velocity field at the spatial point $\vx_{(n)}$. To note is that in \eqref{170830:2045} {\it all} velocity correlations get scaled by the same factor $e^{a_s}$. It is exactly this property of universal scaling that leads for $a_s \neq 0$ to an inconsistency, making the symmetry \eqref{170830:2045} thus unphysical. The issue here is that this transformation is admitted as a {\it statistical} symmetry by the ensemble averaged Navier-Stokes equations \citep{Oberlack10}, a set of statistical equations which dynamically emerge from the deterministic Navier-Stokes equations due to their spatially nonlocal and temporally chaotic behavior. Hence there must exist on the deterministic (fluctuating) level some transformation $\Lambda^\delta_{X_*}$ (which itself need not to be a symmetry) such that on the statistical (averaged) level it emerges as the existing symmetry transformation \eqref{170830:2045}, i.e., such that
$\L\Lambda^\delta_{X_*}\R=\text{S}^\delta_{X_*}$. This is the result of the causality principle, namely that for the effect $\text{S}^\delta_{X_*}$~\eqref{170830:2045} there should exist a cause $\Lambda^\delta_{X_*}$, where obviously the cause itself need not be a symmetry in order to induce a symmetry as an effect. In the following we will prove that for $a_s\neq 0$ such a cause $\Lambda^\delta_{X_*}$ cannot be constructed, thus showing that for $\text{S}^\delta_{X_*}$~\eqref{170830:2045} the classical principle of cause and effect is violated. In other words, the statistical symmetry $\text{S}^\delta_{X_*}$~\eqref{170830:2045} is inconsistent to its underlying deterministic theory, and can only be restored if $a_s=0$, i.e., if the induced symmetry \eqref{170830:2045} and its defining symmetry \eqref{170901:2134} both get broken --- a result already established independently in~Sec.~\ref{Sec1}.

{\bf Statement 1.} For $a_s\neq 0$ the statistical symmetry transformation $\text{S}^\delta_{X_*}$ \eqref{170830:2045} violates the classical principle of causality and thus constitutes an unphysical symmetry {\it per se}.

{\it Proof.} The proof is done by contradiction, initially assuming that $\text{S}^\delta_{X_*}$ \eqref{170830:2045} is {\it not} violating the principle of cause and effect. We start this proof by looking at the structure of transformation $\text{S}^\delta_{X_*}$ \eqref{170830:2045} in its lowest order, which is the transformation of the mean (ensemble averaged) velocity field at every point~$\vx_{(i)}$
\begin{equation}
\L U^k_{(i)}\R^*= e^{a_s}\,\L U^k_{(i)}\R, \;\; \text{for $1\leq i\leq n$, and $1\leq k\leq 3$},\label{170831:0853}
\end{equation}
where the lower index in brackets denotes the particular spatial point to be considered within a multi-point setting, and the upper index the components of the velocity field. Since we assume that the statistical change \eqref{170831:0853} has a deterministic cause, there must exist a transformation on the instantaneous level $\vU_{(i)}\rightarrow \widetilde{\vU}_{(i)}$ such that
\begin{equation}
\L \widetilde{U}^{k}_{(i)}\R=\L U^k_{(i)}\R^*.\label{170831:1122}
\end{equation}
For the specific (trivial) structure of \eqref{170831:0853} in view of all higher-order correlations \eqref{170830:2045} this can only be realized by a componentwise, scalar multiplicative transformation rule $\Lambda^\delta_{X_*}$ of the form
\begin{equation}
\widetilde{U}^{k}_{(i)}=\lambda_{(i)}^k\cdot  U^k_{(i)},\label{170831:1144}
\end{equation}
where $\vlambda_{(i)}=\vlambda(\vx_{(i)},t)$ is in general a continuous random field variable,\footnote[2]{If we choose the scaling factor $\lambda_{(i)}^k$ constant (non-random) for all $i$ and $k$, then the proof continues
as given in \cite{Frewer14.1, Frewer14.2}. The concept of a random dilation as \eqref{170831:1144} and its potential relevance for turbulent scaling was first introduced and discussed in \cite{She15}; see also \cite{She17}.
To note is that if $\lambda_{(i)}^k$ in \eqref{170831:1144} is chosen to be random, then it should be a variable that is continuous, and not a discrete one. Only a continuous random field $\lambda_{(i)}^k$ that is not too rough, will ensure compatibility of the mapped field $\widetilde{U}^{k}_{(i)}$ \eqref{170831:1144} with the Euler or Navier-Stokes equations, to be at least Hölder continuous.}
statistically independent of the velocity field variable $\vU_{(i)}=\vU(\vx_{(i)},t)$, such that
\begin{equation}
\L \widetilde{U}^{k}_{(i)}\R =\L \lambda_{(i)}^k\cdot U^{k}_{(i)}\R=\L \lambda_{(i)}^k\R\cdot \L U^{k}_{(i)}\R=e^{a_s}\cdot \L U^{k}_{(i)}\R=\L U^k_{(i)}\R^*,\;\;\: \forall i, k.\label{170831:1508}
\end{equation}
To note here is the key property of statistical independence for the random scaling factor $\lambda_{(i)}^k$, otherwise we would face different ensemble averaged Navier-Stokes equations for $\L \widetilde{U}^{k}_{(i)}\R$ than for $\L U^{k}_{(i)}\R^*$, which would be in conflict with the equivalence \eqref{170831:1122} and thus of the initial assumption that there exists a cause $\widetilde{U}^{k}_{(i)}$ for the effect
$\L U^{k}_{(i)}\R^*$.\footnote[2]{Caution has to be exercised when applying random transformations as \eqref{170831:1144}. The randomness of $\lambda_{(i)}^k$ is of a different origin and nature than the randomness of $U_{(i)}^k$. Only if the system's field variables (and not the space-time coordinates) get randomly transformed, and only if it occurs statistically independent to the variables it transforms, no peculiar difficulties within a standard statistical analysis of turbulence will arise. If however one of these conditions is not met, one has to be aware that the structure of the statistical field equations will be different from the usual textbook equations. For more details on such random transformations, see e.g. \cite{Filipiak92,McComb14}, where in particular the peculiarities and difficulties of the random Galilean transformations are discussed, explicitly showing that random transformations are ensemble type of operations and not kinematical operations.}

Now since the two-point velocity moments of $\text{S}^\delta_{X_*}$ \eqref{170830:2045} change as
\begin{equation}
\L U^k_{(i)}U^l_{(j)}\R^*= e^{a_s}\,\L U^k_{(i)}U^l_{(j)}\R, \;\; \text{for all $1\leq (i,j)\leq n$, and $1\leq (k,l)\leq 3$},\label{170831:1306}
\end{equation}
which will be caused by
\begin{equation}
\L \widetilde{U}^{k}_{(i)}\widetilde{U}^{l}_{(j)}\R =\L \lambda_{(i)}^kU^{k}_{(i)}\cdot \lambda_{(j)}^l U^{l}_{(j)}\R=\L \lambda_{(i)}^k\lambda_{(j)}^l\R\cdot \L U^{k}_{(i)}U^{l}_{(j)}\R=
e^{a_s}\cdot \L U^{k}_{(i)} U^{l}_{(j)}\R=\L U^k_{(i)} U^{l}_{(j)}\R^*,
\end{equation}
it is possible to determine the correlation coefficient (normalized covariance matrix) of the joint random variable $\vlambda=(\lambda_{(i)}^k)$ between its different components
(see e.g. \cite{Zeidler04})
\begin{align}
\rho^{kl}_{(ij)}& =\frac{\Big\L \big(\lambda^k_{(i)}-\L\lambda^k_{(i)}\R\big)\big(\lambda^l_{(j)}-\L\lambda^l_{(j)}\R\big) \Big\R}{\sqrt{\Big\L\big(\lambda^k_{(i)}-\L\lambda^k_{(i)}\R\big)^2\Big\R}
\sqrt{\Big\L\big(\lambda^l_{(j)}-\L\lambda^l_{(j)}\R\big)^2\Big\R}}\nonumber\\[0.5em]
&=\frac{\big\L\lambda^k_{(i)}\lambda^l_{(j)}\big\R-e^{2a_s}}{\sqrt{\big\L\lambda^k_{(i)}\lambda^k_{(i)}\R-e^{2a_s}}\sqrt{\big\L\lambda^l_{(j)}\lambda^l_{(j)}\R-e^{2a_s}}}
=\frac{e^{a_s}-e^{2a_s}}{\sqrt{e^{a_s}-e^{\vphantom{1^1}2a_s}}\sqrt{e^{a_s}-e^{\vphantom{1^1}2a_s}}}=1,\;\;\;\forall i,j,k,l.
\end{align}
This result tells us that all components of the random variable $\vlambda=(\lambda_{(i)}^k)$ are {\it perfectly} positively correlated, i.e., if we know one component, we know all the other. In other words, between all components there exists a deterministic relationship
\begin{equation}
\lambda_{(i)}^k=c_{(ij)}^{kl}\lambda_{(j)}^l,\;\;\;\forall i,j,k,l,
\end{equation}
where $c_{(ij)}^{kl}>0$ are any positive {\it non-random} variables. But due to the result of \eqref{170831:1508}, we have
\begin{equation}
e^{a_s}=\L \lambda_{(i)}^k\R=\L c_{(ij)}^{kl}\lambda_{(j)}^l\R =c_{(ij)}^{kl}\L\lambda_{(j)}^l\R=c_{(ij)}^{kl}e^{a_s},
\end{equation}
that implies
\begin{equation}
c_{(ij)}^{kl}=1, \;\;\;\forall i,j,k,l.
\end{equation}
Hence, the random multi-component transformation rule \eqref{170831:1144} is equivalent to
\begin{equation}
\widetilde{U}^{k}_{(i)}=\lambda\cdot  U^k_{(i)},\;\;\;\forall i,k,\label{170831:1518}
\end{equation}
where $\lambda$ is a scalar continuous random variable statistically independent to all components of the velocity field. In other words, the cause $\Lambda^\delta_{X_*}$ for the effect $\text{S}^\delta_{X_*}$ \eqref{170830:2045} thus reduces from a random multi-component transformation \eqref{170831:1144} to a random scalar transformation \eqref{170831:1518}, which for the general multi-point moment then reads
\begin{multline}
\L \widetilde{\vU}_{(1)}\otimes\cdots\otimes \widetilde{\vU}_{(n)}\R =\L \lambda \vU_{(1)}\otimes\cdots\otimes \lambda \vU_{(n)}\R=\L \lambda^n \R \L \vU_{(1)}\otimes\cdots\otimes \vU_{(n)}\R \\[0.5em]
=e^{a_s}\,\L\vU_{(1)}\otimes\cdots\otimes \vU_{(n)}\R
=\L\vU_{(1)}\otimes\cdots\otimes \vU_{(n)}\R^* ,\hspace{0.8cm}
\end{multline}
with the result
\begin{equation}
\L \lambda^n\R =e^{a_s},\;\;\;\forall n\geq 1.\label{170831:1605}
\end{equation}
Say $p=p(\lambda)$ is the PDF of the continuous random variable $\lambda\neq 0$, then the result \eqref{170831:1605} implies the equation
\begin{equation}
0=\L \lambda^{n+1}\R-\L \lambda^{n}\R=\int d\lambda\, (\lambda^{n+1}-\lambda^{n})\, p(\lambda),\;\;\;\forall n\geq 1,
\label{220430:2030}
\end{equation}
however which, for all $n\geq 1$ and $\lambda\neq 0$, can only be fulfilled if$\,$\footnote[2]{Note that if $\lambda$ would be a discrete random variable taking only the two special integer values 0 and 1, 
then a further solution to \eqref{220430:2030} would be the Bernoulli distribution: $p(\lambda)=e^{a_s}\delta(\lambda-1)+(1-e^{a_s})\delta(\lambda-0)$, with $a_s<0$.\linebreak[4] But such a solution is of course incompatible with the Euler or Navier-Stokes equations, since the {\it instantaneous} fields, i.e.~the full fields and not just their fluctuating part, get mapped by \eqref{170831:1518} to discrete discontinuous fields.}
\begin{equation}
p(\lambda)=\delta(\lambda-1),
\label{220430:2032}
\end{equation}
i.e., when the variable $\lambda$ is {\it not} random but constant\footnote[3]{The combined result that the scaling factor $\lambda_{(i)}^k$ \eqref{170831:1144} has to be isotropic
$\lambda_{(i)}^k=\lambda$ and constant $\L\lambda\R=\lambda$ ultimately validates in retrospect the correctness of the proof assumptions made in \cite{Frewer14.1, Frewer14.2}.} in taking the particular value $\lambda=1$. But this implies again that $e^{a_s}=1$, which for $a_s\neq 0$ leads to a contradiction. Hence, for $a_s\neq 0$ there is no cause for the effect $\text{S}^\delta_{X_*}$ \eqref{170830:2045}, thus violating the principle of cause and effect.
\hfill$\square$

\vspace{1em}
The next step is to prove the general case that the symmetry $\text{T}_{X_*}$~\eqref{170830:1150} is violating the principle of cause and effect for {\it all} admissible translation functions $b_{(n)}^0$, and not only for those specified in \eqref{170901:2143}. Along with the LMN normalization constraint, the general PDF symmetry $\text{T}_{X_{*}}$~\eqref{170830:1150} induces the following symmetry transformation for the velocity correlations \citep{Oberlack15, Oberlack14}
\begin{equation}
\text{S}_{X_*}\!:\;\; t^*=t,\;\; \vx_{(i)}^*=\vx_{(i)},\;\; \L \vU_{(1)}\otimes\cdots\otimes \vU_{(n)}\R^*=e^{a_s}\, \L \vU_{(1)}\otimes\cdots\otimes \vU_{(n)}\R + (1-e^{a_s})\,\vC_{(n)},\;\; \forall n\geq 1,\label{170901:2322}
\end{equation}
where $\vC_{(n)}=\vC_{(n)}(\vx_{(1)},\dotsc,\vx_{(n)},t)$ is any solution of the instantaneous velocity multi-point correlation (MPC) equations \citep{Oberlack10}.

{\bf Statement 2.} For $a_s\neq 0$ the symmetry transformation $\text{S}_{X_*}$ \eqref{170901:2322} violates the classical principle of causality, i.e., there exists no cause $\Lambda_{X_*}$ on the level of the underlying deterministic Navier-Stokes equations of any type and form such that the statistical symmetry $\text{S}_{X_*}$ \eqref{170901:2322} results as an effect, constituting thus an unphysical symmetry {\it per se}.\footnote[4]{Formally this statement says: No $\Lambda_{X_*}$ exists such that $\L \Lambda_{X_*}\R=\text{S}_{X_*}$, where the ansatz for the cause~$\Lambda_{X_*}$ is fully unrestricted and not conditioned to necessarily be a symmetry.}

{\it Proof.}  As before, the proof is done by contradiction, initially assuming that $\text{S}_{X_*}$ \eqref{170901:2322} is {\it not} violating the principle of cause and effect. We start again by looking at the structure of the transformation $\text{S}_{X_*}$ \eqref{170901:2322} in its lowest order
\begin{equation}
\L U^k_{(i)}\R^*= e^{a_s}\,\L U^k_{(i)}\R+ (1-e^{a_s})\,C_{(i)}^k, \;\; \text{for $1\leq i\leq n$, and $1\leq k\leq 3$},\label{170902:0024}
\end{equation}
which again, as before in \eqref{170831:1144}, can only be realized on the instantaneous level by a componentwise, scalar multiplicative cause $\Lambda_{X_*}$, however, now in the more general linear stochastic~form
\begin{equation}
\widetilde{U}^{k}_{(i)}=\lambda_{(i)}^k\cdot  U^k_{(i)}+\gamma_{(i)}^k,\label{170902:0904}
\end{equation}
where $\vlambda_{(i)}=\vlambda(\vx_{(i)},t)$ and $\vgamma_{(i)}=\vgamma(\vx_{(i)},t)$ are two continuous random field variables, both statistically independent of the velocity
field variable $\vU_{(i)}=\vU(\vx_{(i)},t)$, such that
\begin{equation}
\L \widetilde{U}^{k}_{(i)}\R =\L \lambda_{(i)}^k U^{k}_{(i)}+\gamma_{(i)}^k\R=\L \lambda_{(i)}^k\R  \L U^{k}_{(i)}\R+\L \gamma_{(i)}^k\R=e^{a_s} \L U^{k}_{(i)}\R
+(1-e^{a_s})\,C_{(i)}^k=\L U^k_{(i)}\R^*,\;\;\: \forall i, k.\label{170902:0905}
\end{equation}
Note that no assumptions are made on the statistical dependence between $\vlambda_{(i)}$ and $\vgamma_{(i)}$, i.e., whether $\L \vlambda_{(i)}\otimes\vgamma_{(j)}\R = \L\vlambda_{(i)}\R\otimes\L\vgamma_{(j)}\R$ is assumed or not is irrelevant at this point.
Now, since the two-point velocity moments of $\text{S}_{X_*}$ \eqref{170901:2322} transform as
\begin{equation}
\L U^k_{(i)}U^l_{(j)}\R^*= e^{a_s}\,\L U^k_{(i)}U^l_{(j)}\R+(1-e^{a_s})\,C_{(ij)}^{kl}, \;\; \text{for all $1\leq (i,j)\leq n$, and $1\leq (k,l)\leq 3$},\label{170831:1306}
\end{equation}
which will be caused by
\begin{align}
\L \widetilde{U}^{k}_{(i)}\widetilde{U}^{l}_{(j)}\R &=\big\L (\lambda_{(i)}^kU^{k}_{(i)}+\gamma_{(i)}^k)(\lambda_{(j)}^l U^{l}_{(j)}+\gamma_{(j)}^l)\big\R\nonumber\\[0.5em]
&=\L \lambda_{(i)}^k\lambda_{(j)}^l\R\L U^{k}_{(i)}U^{l}_{(j)}\R +\L \lambda_{(i)}^k\gamma_{(j)}^l\R\L U^{k}_{(i)}\R +\L \gamma_{(i)}^k\lambda_{(j)}^l\R\L U^{l}_{(j)}\R
+\L \gamma_{(i)}^k\gamma_{(j)}^l\R\qquad
\nonumber\\[0.5em]
&=e^{a_s}\L U^{k}_{(i)} U^{l}_{(j)}\R+(1-e^{a_s})\,C_{(ij)}^{kl}=\L U^k_{(i)} U^{l}_{(j)}\R^*,
\end{align}
the following statistical relations are obtained
\begin{equation}
\L \lambda_{(i)}^k\lambda_{(j)}^l\R=e^{a_s},\qquad \L \lambda_{(i)}^k\gamma_{(j)}^l\R=0, \qquad \L \gamma_{(i)}^k\gamma_{(j)}^l\R= (1-e^{a_s})\,C_{(ij)}^{kl},\qquad\forall i,j,k,l.\label{170902:1102}
\end{equation}
Hence, for the random variable $\lambda_{(i)}^k$ the same result as in the previous proof is obtained: Since the correlation $\L \lambda_{(i)}^k\lambda_{(j)}^l\R$ is again isotropic, the multi-component random variable $\lambda_{(i)}^k$ can be equivalently identified as the scalar random variable $\lambda=\lambda_{(i)}^k$, and since the general transformation $\text{S}_{X_*}$ \eqref{170901:2322} dictates $\lambda$ to correlate
again as $\L \lambda^n\R= e^{a_s}$ \eqref{170831:1605}, for all $n\geq 1$, the same result is obtained again, namely that $\lambda$ is a non-random variable with the constant value $\lambda=1$. Since this implies $a_s=0$, i.e., the breaking of the symmetry $\text{S}_{X_*}$ \eqref{170901:2322} down to an identity transformation $\L \vU_{(1)}\otimes\cdots\otimes \vU_{(n)}\R^*=\L \vU_{(1)}\otimes\cdots\otimes \vU_{(n)}\R$, there simply exists no cause on the level of the underlying deterministic Navier-Stokes equations such that the statistical symmetry transformation $\text{S}_{X_*}$ \eqref{170901:2322} can result as an effect for $a_s\neq 0$. \hfill$\square$

\vspace{1em} What is left to prove is that also the statistical translation symmetry $\text{T}_{X_{**}}$~\eqref{170901:1745} itself is unphysical {\it per se}. As before, this will again be done on the level of the multi-point moments, since the proof on the level of the PDF's has already been generally given in \cite{Frewer15,Frewer14.1}. Since the ingredients of this proof are the same as in the two previous ones, it will not be carried out in such detail as before; only the key idea will be given. The induced symmetry for the velocity correlations of the general PDF symmetry $\text{T}_{X_{**}}$~\eqref{170901:1745} (along with the LMN normalization constraint) is obviously given by \citep{Oberlack15, Oberlack14}
\begin{equation}
\text{S}_{X_{**}}\!:\;\; t^*=t,\;\; \vx_{(i)}^*=\vx_{(i)},\;\; \L \vU_{(1)}\otimes\cdots\otimes \vU_{(n)}\R^*=\L \vU_{(1)}\otimes\cdots\otimes \vU_{(n)}\R + \vC_{(n)},\;\;\forall n\geq 1,\label{170902:1217}
\end{equation}
where $\vC_{(n)}=\vC_{(n)}(\vx_{(1)},\dotsc,\vx_{(n)},t)$ is again any solution of the instantaneous velocity multi-point correlation (MPC) equations \citep{Oberlack10}.

{\bf Statement 3.} For $\vC_{(n)}\neq \mathbf{0}$ the symmetry transformation $\text{S}_{X_{**}}$ \eqref{170902:1217} violates the classical principle of causality, i.e., there exists no cause $\Lambda_{X_{**}}$ on the level of the underlying deterministic Navier-Stokes equations of any type and form such that the statistical symmetry $\text{S}_{X_{**}}$ \eqref{170902:1217} results as an effect, constituting thus, next to $\text{S}_{X_*}$ \eqref{170901:2322}, also an unphysical symmetry {\it per se}.

{\it Proof.} The proof is done again by contradiction. Assuming that the statistical effect $\text{S}_{X_{**}}$~\eqref{170902:1217} has a deterministic cause, then its lowest correlation order
\begin{equation}
\L U^k_{(i)}\R^*= \L U^k_{(i)}\R+C_{(i)}^k, \;\; \text{for $1\leq i\leq n$, and $1\leq k\leq 3$},\label{170902:1350}
\end{equation}
can in view of all higher-order correlations \eqref{170902:1217} only be caused by
\begin{equation}
\widetilde{U}^{k}_{(i)}=U^k_{(i)}+\gamma_{(i)}^k,\label{170902:1353}
\end{equation}
where $\vgamma_{(i)}=\vgamma(\vx_{(i)},t)$ is a continuous random field variable, statistically independent of the velocity
field variable $\vU_{(i)}=\vU(\vx_{(i)},t)$, such that
\begin{equation}
\L \widetilde{U}^{k}_{(i)}\R =\L U^{k}_{(i)}+\gamma_{(i)}^k\R=\L U^{k}_{(i)}\R+\L \gamma_{(i)}^k\R=\L U^{k}_{(i)}\R
+C_{(i)}^k=\L U^k_{(i)}\R^*,\;\;\: \forall i, k.\label{170902:1354}
\end{equation}
The next higher order of $\text{S}_{X_{**}}$ \eqref{170902:1217} will then be caused by the transformation
\begin{align}
\L \widetilde{U}^{k}_{(i)}\widetilde{U}^{l}_{(j)}\R =\big\L (U^{k}_{(i)}+\gamma_{(i)}^k)(U^{l}_{(j)}+\gamma_{(j)}^l)\big\R
&=\L U^{k}_{(i)}U^{l}_{(j)}\R +\L \gamma_{(i)}^k\R\L U^{l}_{(j)}\R+\L \gamma_{(j)}^l\R\L U^{k}_{(i)}\R
+\L \gamma_{(i)}^k\gamma_{(j)}^l\R\hspace{1cm}
\nonumber\\[0.5em]
&=\L U^{k}_{(i)} U^{l}_{(j)}\R+C_{(ij)}^{kl}=\L U^k_{(i)} U^{l}_{(j)}\R^*,
\end{align}
to give the statistical relations up to second order as
\begin{equation}
\L \gamma_{(i)}^k\R=0, \qquad \L \gamma_{(i)}^k\gamma_{(j)}^l\R=C_{(ij)}^{kl},\qquad\forall i,j,k,l.\label{170902:1410}
\end{equation}
But when taking the result of the first order transformation \eqref{170902:1354}, the above second order result simply implies
\begin{equation}
0=\L\gamma_{(i)}^k\R=C_{(i)}^k, \;\;\;\forall i,k.
\end{equation}
This process can now be continued to arbitrary high order, where in each case the result of the higher $(n+1)$-th order will imply that the lower $n$-th order correlation
$C_{(i_1\dotsc i_n)}^{k_1\dotsc k_n}=\L \gamma_{(i_1)}^{k_1}\cdots\gamma_{(i_n)}^{k_n}\R$ has to be zero. Hence, for $\vC_{(n)}\neq \mathbf{0}$ no cause on the level of the underlying deterministic Navier-Stokes equations of any type and form exists such that the statistical translation symmetry $\text{S}_{X_{**}}$~\eqref{170902:1217} can result as an effect. \hfill$\square$

\vspace{1em} {\bf Summary.} The ``new" statistical symmetries $\text{T}_{X_*}$~\eqref{170830:1150} and $\text{T}_{X_{**}}$~\eqref{170901:1745}, its correspondingly induced symmetries $\text{S}_{X_*}$ \eqref{170901:2322}
and  $\text{S}_{X_{**}}$~\eqref{170902:1217}, as well as any of its variants, are all physically spurious symmetries, not only in \cite{Oberlack17}, but also in \cite{Oberlack10}, \cite{Oberlack14.1}, \cite{Oberlack14} and \cite{Oberlack15}, simply due to the fact of not providing the necessary statistical link to the underlying (fine-grained) Navier-Stokes equations in their performed symmetry analysis for the (coarse-grained) LMN or MPC equations. Including this link \citep{Frewer15,Frewer14.1,Frewer14.2} will break all these symmetries and thus will not give rise to such unphysical symmetries in the first place, showing that in turbulence it is necessary to reveal all information available.

Lastly, in this context it is to be noted that the (trivial) PDF symmetries mentioned in \cite{Kozlov12,Kozlov13} and referred to in \cite{Oberlack17} originating solely from the linearity of the PDF-describing Fokker-Planck equation are spurious symmetries as well. They are just mathematical artefacts with no cause in its underlying or corresponding stochastic differential equation (SDE), if set to be non-linear. In contrast, of course, to the determined (non-trivial) $Y$-symmetries of the Fokker-Planck equation, which, although they are not related to any symmetries of the SDE, are still caused by the SDE, not as symmetries, but as usual yet non-trivial space-time transformations. For a general discussion on this subtle issue, please see, e.g.,~\cite{Frewer16.2}.

\section{Conclusion and final remarks \label{Sec4}}

Considering the facts established in Secs.~\ref{Sec1}-\ref{Sec3}, the only symmetries left for further analysis are the classical symmetries $X_1$-$X_{12}$ (Eqs.~[41-46]) of the Navier-Stokes equations when formulated statistically in the framework of the LMN equations. But this is not new and also not a surprising result, since this is to be expected.
The two ``new" infinite-dimensional groups $X_*$ (Eq.~[47]) and $X_{**}$ (Eq.~[48]), however, are broken and thus not available to serve the unrealistic expectation in \cite{Oberlack17} that a ``non-Gaussian solution for the tails of pdf's could be derived based on the symmetries of the corresponding LMN equations for velocity differences" (p.~[11]). This expectation, which in their present study is directly linked to the expectation in trying to grasp the ``phenomenon of internal intermittency of turbulence" (p.~[11]) with the help of symmetries, is unrealistic in so far as intermittency has the well-known property to rather break than to restore symmetries, not only on the fine grained (fluctuating) but also on the coarse-grained (averaged) level. And since the available symmetries $X_1$-$X_{12}$ (excluding the broken and unphysical symmetries $X_{*}$ and $X_{**}$ from this set due to the facts established in this Comment) collectively only form a global finite and not a local infinite-dimensional Lie-algebra, they are more prone to be broken in a turbulent and intermittent flow, thus favoring the situation of symmetry breaking, where inside this finite algebra a noncompact group as that of global scale invariance is again more prone to be broken than a compact group as that of global rotation invariance
\citep{Benzi10,Grauer12,Friedrich16,Iyer17}.

\appendix

\section{Outline of a causal structure induced by the Navier-Stokes equations\label{SecA}}

This section will briefly outline the particular causal structure we address here in this Comment within the context of trying to classify the principle of causality in general, which itself is a difficult undertaking and definitely beyond the scope of this article.

To grasp the phenomenon ``cause and effect" in all its facets is not easy and straightforward. In particular, since it can be formulated in many different ways depending on the type of physics one is doing: There are classical, relativistic, quantum mechanical, thermodynamical and cosmological formulations for it (see e.g. \cite{Beebee09}). It still is a philosophical question, whether the concept of cause and effect should or can be posed as a fundamental principle of nature \citep{Frisch14,Pearl09,Woodward03,Dowe00,Salmon98}. Maybe a general formulation for all spatial and temporal scales cannot be formulated, since its validity in the microscopic, mesoscopic and macroscopic world is different.

For example in the microscopic world, quantum mechanics shows that some events can happen without a cause, like radioactive decay. Also the underlying dynamical equations themselves describing quantum mechanical processes, like e.g. the Schrödinger equation, have no cause. Yet this does not mean that quantum mechanics as a physical theory is void of causation. Rather than the quantum dynamical equation itself, it is the time evolution of a quantum state according to this equation and its response to operators where causality in quantum mechanics is found. In other words, causality is seen in the development of the probabilities of a systems' experimental outcomes, but not in the process when individual events are explicitly measured, which seem to be random. Ever since the first mathematical developments of quantum theory nearly a century ago, this phenomenon is known as the ``measurement problem" famously formulated e.g. by \cite{Neumann32} as ``a [quantum] state ... under the action of an energy operator ... is purely causal," while, ``on the other hand, the state ... which may measure a [given] quantity ... undergoes in a measurement a non-causal change".\footnote[2]{For a critical examination of this statement by von Neumann, see e.g. \cite{Plotnitsky09}. For a more broad critical review of von Neumann's understanding of causality and determinism in the context of his hidden-variables-theorem, see \cite{Redei01}, pp.~173-188.} On the microscopic level cause and effect can even be violated within the uncertainty principle, experienced as violation of conservation laws as that of energy on very short timescales. Things that are strict laws in the macroscopic world, such as the conversation of mass and energy, can be broken in the quantum~world.

On the other hand, quantum mechanics also shows that the classical concept of cause and effect survives as a consequence of the collective behaviour of large quantum systems. However, in the macroscopic world on very large (cosmological) scales one faces new problems with the concept of cause and effect. Does the origin of the universe have a cause? What is the cause for the existence of an inertial frame if it exists in a strict (non approximate) sense; is it according to Mach's principle that all mass of the universe determines the structure and behavior of an inertial system and thus can be distinguished from all other (accelerated) frames?\footnote[3]{This discussion was first introduced by \cite{Mach83} and is still an open discussion today \citep{Friedman83,Barbour95,Barbour04,Penrose05,Pfister15}.\\ \phantom{x}} A question that directly relates to: What is the cause for Newton's first law of motion?

However, the issue we deal here with in this Comment is very much simpler. First of all, we refer to the classical (common sense) principle of causality which in its most general version can be formulated as: Everything that came into being (``effect") has a cause, where (i) the cause always exists or comes {\it before} the effect (asymmetry), and (ii) where the {\it same} cause always gives the {\it same} effect (determinism). Secondly, we consider a deterministic classical equation as our starting point, namely the Navier-Stokes (momentum) equation which is based on Newton's second law in a continuum, to be symbolically denoted in the following by $\mathcal{N}=0$. For this equation, two independent, exogenic common sense causations can be formulated:\pagebreak[4]
\begin{itemize}
\item[(1)]
If $F$ is an {\it external} force added to $\mathcal{N}=0$, then $F$ is a cause for the motion of the fluid's velocity~field.
\item[(2)]
If $\mathcal{N}=0$, due to its unstable behaviour leading up to chaotic solutions, is ensemble averaged over these solutions, then the corresponding statistically formulated Navier-Stokes equation $\L\mathcal{N}\R=0$ (or any other statistical set of equations, like the LMN system of PDF equations) is the effect of the instability emerging from the deterministic Navier-Stokes equation $\mathcal{N}=0$. Hence, since the deterministic Navier-Stokes equation implies its statistical equations and not opposite, a strict principle of cause and effect is formulated by this~asymmetric relation.
\end{itemize}
It is the latter causation we consider here in our Comment in Sec.~\ref{Sec3}. Instead of focussing on the equations, the above formulation can also be equivalently reformulated with the focus on the solutions of these equations. To obtain this reformulation, it will be based on a practical day-by-day example: Assume we have performed a DNS for a particular flow configuration imposed on~$\mathcal{N}=0$. As a result we obtain a raw ensemble data set, let's call it~$\mathcal{D}$ and say it's of size of a few gigabyte (GB). This is our cause, which, up to the numerical resolution considered, represents a solution of the simulated equation, to be denoted by $\mathcal{N}_\mathcal{D}=0$. When averaging this set up to a certain statistical order, to get the resulting data set $\L \mathcal{D}\R$, say of size of a few hundred kilobyte (kB), we obtain an effect, representing a solution then of the corresponding statistical equation $\L\mathcal{N}\R_{\L \mathcal{D}\R}=0$.
Hence, since $\L \mathcal{D}\R$ results from $\mathcal{D}$ and not opposite, a strict chain of cause and effect can again be formulated, equivalent to that of (2), however, now for the solutions~of~$\mathcal{N}=0$:
\begin{itemize}
\item[($2^{\,\prime}$)] If $\mathcal{D}$ is an ensemble set of solutions of $\mathcal{N}=0$, then $\mathcal{D}$ is the cause for the statistical solution~$\L \mathcal{D}\R$~of~$\L\mathcal{N}\R=0$.
\end{itemize}
Now, imagine we intervene and manipulate the raw GB-data set $\mathcal{D}$ by transforming, e.g., the instantaneous velocity fields $\vU$ to some new values $\widetilde{\vU}$, to get the new raw GB-data set $\widetilde{\mathcal{D}}$. Then two things can happen when averaging this set: Either $(a)$ the transformations on the fine-grained GB-level cancel in average, and we obtain an invariant result on the coarse-grained kB-level, i.e., on the averaged level we get the same data set as before $\L\widetilde{\mathcal{D}}\R=\L\mathcal{D}\R$, or $(b)$ the transformations on the fine-grained GB-level do not cancel in average, and we obtain a different result on the coarse-grained kB-level, i.e., on the averaged level we get a different data set as before $\L\widetilde{\mathcal{D}}\R=:\L \mathcal{D}\R^*$, where the new (transformed) set $\L D\R^*$ is not equal to the initial (untransformed) one,~i.e.,~where~$\L \mathcal{D}\R^*\neq \L\mathcal{D}\R$.

In the first case $(a)$, we don't see any change on the averaged (coarse-grained) kB-level, although an active process is happening on the instantaneous (fine-grained) GB-level in going from $\mathcal{D}$ to $\widetilde{\mathcal{D}}$. In the second case $(b)$, however, we clearly observe a change on the averaged (coarse-grained) kB-level in that $\L\mathcal{D}\R$ goes to $\L \mathcal{D}\R^*$, a change or an effect that clearly has a cause on the instantaneous (fine-grained) GB-level, namely in that $\mathcal{D}$ goes to $\widetilde{\mathcal{D}}$. Hence, we can conclude that if there is a change on the averaged level, then there must be a cause for it on the instantaneous level, which to construct (if the cause is not explicitly given beforehand) can be in general a very difficult task since it is an inverse problem. Surely, the opposite we may not conclude, as the first case $(a)$ clearly shows, i.e., if no changes on the averaged level happen, then we may not conclude that the instantaneous level stays unchanged too.

Now, the two new symmetries Eqs.~[47-48] in \cite{Oberlack17} are pure statistical transformations that clearly change the statistical fields from $f_n$ to $f_n^*$ (see \eqref{170830:1150} and \eqref{170901:1745}, respectively), and hence, in each case, this change should have a cause on the instantaneous level, where, of course, the cause itself need not to be a symmetry. Furthermore, since these two statistical transformations \eqref{170830:1150} and \eqref{170901:1745} are so simple and trivial, it is very easy to find a unique cause for the change of the lowest velocity moment, the mean velocity (see e.g.~\eqref{170831:1508}), but, as our proof in Sec.~\ref{Sec3} shows, is not compatible to any of the higher order velocity moments. Hence, no overall cause on the instantaneous GB-level can be constructed for the change that happens\pagebreak[4] on the averaged kB-level. Important to note here is that both statistical transformations \eqref{170830:1150} and \eqref{170901:1745} are admitted as symmetries by the statistical equations but not by the statistical data itself, which does not stay invariant since it changes from $\L\mathcal{D}\R$ to a different~$\L\mathcal{D}\R^*$, and exactly this change should have a cause on the instantaneous (fine-grained) level.

The deeper reason why no cause for the statistical changes \eqref{170830:1150} and \eqref{170901:1745} can be found or constructed is that Wac{\l}awczyk~{\it et~al.} are not providing to their symmetry analysis the essential information that any $(n\!>\!2)$-point velocity moment is a true non-linear object. Providing this information would not allow for such unphysical symmetries as \eqref{170830:1150} and \eqref{170901:1745} in the first place. This can also be independently seen with the issue of the LMN separation constraint in Sec.~\ref{Sec1}. Why are Wac{\l}awczyk's~{\it et~al.} new (linearly induced) symmetries \eqref{170830:1150} and \eqref{170901:1745} incompatible to this internal constraint? Because this constraint \eqref{170830:1149} is non-linear!

To close this section, it is worth to mention that a successful statistical theory of Navier-Stokes turbulence can only be brought about if all information of the theory is revealed, including its causal link to the underlying deterministic Navier-Stokes equation. It is incorrect to think that all information about a system can be represented only in terms of statistical correlations among its variables (see e.g. \cite{Frisch14}). One of the key reasons is that correlation does not imply causation, as for instance correctly pointed out by \cite{Pienaar17}: ``The concept of causal information goes beyond that of correlation ... [it]~tells us more ... [and] is different from correlations because it tells us how the system changes under interventions." Hence, next to correlations we also need to consider causal knowledge in turbulence as put forward in this~Comment.

\bibliographystyle{jfm}
\bibliography{References}

\end{document}